# Redshift and the Rotating Gravitational Field

Walter J. Christensen Jr.
Cal Poly Pomona University
3801 W. Temple Ave
Pomona CA 91768

wjchristense@csupomona.edu

July, 26 2010

**Abstract**

Previously it was shown that if a weak gravitational field is modeled as a background of oscillating gravitons described by normal coordinates, then the field naturally exhibits rotational kinetic energy[1]. The conformal metric associated with this oscillatory motion is given by $g_{\mu\nu} = e^{i\omega t}\eta_{\mu\nu}$, and the corresponding energy momentum tensor by $T_{\mu\nu} = \frac{1}{2} I\omega^2 \eta_{\mu\nu}$. In this paper the metric is extended to include constant radiant energy thereby amending the spacetime metric to: $g_{\mu\nu} = e^{i(\omega+\rho)t}\eta_{\mu\nu}$. The energy momentum tensor then becomes: $T_{\mu\nu} = [\frac{1}{2} I\omega^2 + (3c^2/16G)(\rho^2 + 2n\omega\rho)]$. Analyzing this energy equation at the microscopic level, where energies are assumed to become discrete, it is found a photon of frequency $\nu_0$ traversing through a rotating gravitational field (having frequency $\nu_g$) becomes coupled to the field and redshifted by the amount $\nu' = \nu_0 - r(\nu_0 \nu_g)^{1/2}$.

## I. Introduction

Fundamental to the theory of general relativity is the coupling existing between the gravitational field and the energy-momentum source $T_{\mu\nu}$; if one changes, so too, will the other. In particular, if the gravitational field undergoes oscillations then there must be a causal source inducing these oscillations. If so, this suggests the gravitational system can be treated like a coupled spring and driver. Though coupled motion can be quite complex, not even periodic, it can always be described in terms of a set of normal coordinates having the property that each coordinate oscillates with a single, well-defined frequency with no coupling among them[2].

The goal then will be to describe the motion of the energy-momentum-source $T_{\mu\nu}$, simply by knowing that the gravitational field is oscillating. This can be accomplished analogously through the classical approach of analyzing small displacements about a point of equilibrium and then solving for the normal coordinates, a procedure that is also well known in gravitational literature[3].

Once these coordinates are identified they can be brought into the language of general relativity. The metric tensor is then constructed and energy-momentum tensor calculated from the Einstein Tensor $G_{\mu\nu}$. The resulting diagonal tensor $T_{\mu\nu}$ will be shown to have components of pure rotational kinetic energy density. In classical physics this diagonal-kinetic-energy result is a necessary condition imposed by normal coordinates. Therefore the method presented here of extending normal coordinates into general relativity is promising. Furthermore, though the diagonal rank-two tensor $T_{\mu\nu}$ is shown to be constant and real, the contravariant tensor $T^{\mu\nu}$ necessarily turns out to be complex. However, $T^{\mu\nu}$ becomes real and equal to $T_{\mu\nu}$ every one-fourth the period of the fundamental mode of oscillation of the normal coordinates. Countable rotational symmetry, together with Noether's theorem[4], suggests the energy-momentum tensor $T^{\mu\nu}$ is conserved. Finally, radiant energy is introduced into the metric and the resulting energy momentum tensor suggests that photons entering a rotational gravitational field become redshifted and greater so over distance.

## II. Normal Coordinates

Einstein's gravitational field equations express a causal link between the energy-momentum-source $T_{\mu\nu}$, and spacetime curvature associated with the tensor $G_{\mu\nu}$. The purpose of this paper will be to determine the causal motion of the source inducing oscillations in the gravitational field. The problem is reminiscent of a classical oscillator and driver and will be our starting point. We begin with a Lagrangian representing small oscillations about a point of equilibrium.

$$L = \frac{1}{2}\left(T_{ij}\dot{\eta}_i\dot{\eta}_j - V_{ij}\eta_i\eta_j\right) \tag{2.1}$$

The $\eta_i$'s represent small deviations from the generalized coordinates $q_{0i}$, such that $q_i = q_{0i} + \eta_i$. Classically the $\eta$'s subsequently become the generalized coordinates for the equations of motion, wherein the kinetic energy has diagonal components only.

$$T_i\ddot{\eta}_i - V_{ij}\eta_j = 0 \qquad \text{(no sum over i)} \tag{2.2}$$

The solution[5] to (2.2) has the normal coordinate form of:

$$\eta_i = C_\kappa e^{-i\omega_\kappa t} \tag{2.3}$$

Assuming these coordinates quasi-describe oscillations in the gravitational field, by the principle of equivalence let a general relativistic coordinate basis $e_\mu$ experience the accelerations expressed by (2.3). Furthermore let the coefficients of (2.3) be set equal to one and let negative one-half be introduced in front of the angular velocity. These small changes allow for the motion of the energy-momentum source to become more apparent. Rayleigh's principle[6, 7] is applied, and the coordinate frequencies $\omega_\kappa$ reduce to the fundamental mode of oscillation, $\omega$, having the greatest intensity. The average kinetic energy $\langle T \rangle$ is then equal to the average potential energy $\langle U \rangle$.

With the preceding adjustments made, the basis for the general relativistic coordinate system is constructed from the modified normal coordinates[8, 9]:

$$(e_\mu)_\nu \equiv e^{\frac{i\omega t}{2}} \delta_{\mu\nu} \tag{2.4}$$

By definition the inner product of any two such basis elements $e_\mu, e_\nu$ yields the metric tensor.

$$g_{\mu\nu} \equiv e_\mu \cdot e_\nu = e^{i\omega t} \delta^\mu{}_\nu = e^{i\omega t} \eta_{\mu\nu} \tag{2.5}$$

As with the mechanical oscillation problems it is understood that only the real part of this complex metric corresponds to physical measurement.

### III. Energy-Momentum Tensor

The constructed weak field metric $g_{\mu\nu}$ is applied to Einstein tensor $G_{\mu\nu}$. A straightforward calculation produces the energy-momentum tensor $T_{\mu\nu}$[10], and together they form a linearized theory of gravitation[11]:

$$T_{\mu\nu} = \frac{c^2}{16\pi G} \begin{bmatrix} -\frac{3\omega^2}{2} & 0 & 0 & 0 \\ 0 & \frac{\omega^2}{2} & 0 & 0 \\ 0 & 0 & \frac{\omega^2}{2} & 0 \\ 0 & 0 & 0 & \frac{\omega^2}{2} \end{bmatrix} \tag{3.1}$$

$T_{\mu\nu}$ is separated out to show its rotational kinetic energy density form.

$$T_{\mu\nu} = \frac{1}{2}\frac{c^2}{8\pi G}\begin{bmatrix} -\frac{3}{2} & 0 & 0 & 0 \\ 0 & \frac{1}{2} & 0 & 0 \\ 0 & 0 & \frac{1}{2} & 0 \\ 0 & 0 & 0 & \frac{1}{2} \end{bmatrix}\begin{bmatrix} \omega^2 & 0 & 0 & 0 \\ 0 & \omega^2 & 0 & 0 \\ 0 & 0 & \omega^2 & 0 \\ 0 & 0 & 0 & \omega^2 \end{bmatrix} \quad (3.2)$$

The moment of inertia and angular frequency matrices are defined to be:

$$I \equiv \frac{c^2}{8\pi G}\begin{bmatrix} -\frac{3}{2} & 0 & 0 & 0 \\ 0 & \frac{1}{2} & 0 & 0 \\ 0 & 0 & \frac{1}{2} & 0 \\ 0 & 0 & 0 & \frac{1}{2} \end{bmatrix}; \quad \tilde{\omega}^2 \equiv \begin{bmatrix} \omega^2 & 0 & 0 & 0 \\ 0 & \omega^2 & 0 & 0 \\ 0 & 0 & \omega^2 & 0 \\ 0 & 0 & 0 & \omega^2 \end{bmatrix} \quad (3.3)$$

In compact tensor notation the energy-momentum tensor becomes

$$T_{\mu\nu} \equiv \frac{1}{2}I\tilde{\omega}^2 \quad (3.4)$$

If the angular velocity is replaced by $\frac{v}{r} = \omega$, then equation (3.1) resembles an energy-momentum tensor for a radiation dominated perfect fluid[12]--in particular a perfect fluid of gravitons. It is important to realize $T_{\mu\nu}$ was derived from Einstein's gravitational wave equation based on a variational principle, and not upon the prejudice of definition. Furthermore, it is interesting to observe that, although, $T_{\mu\nu}$ is completely real, its contravariant counterpart $T^{\mu\nu}$ is necessarily complex.

$$T^{\mu\nu} = g^{\mu\alpha}g^{\nu\beta}T_{\alpha\beta} = e^{-2i\omega t}\eta^{\mu\alpha}\eta^{\nu\beta}T_{\alpha\beta} \quad (3.5)$$

This result shows the Einstein tensor $G_{\mu\nu}$ and its metric constructed from normal coordinates, are able to separate the energy momentum-tensor $T^{\mu\nu}$ into real and imaginary parts through a time rotation. Although time-wise there are uncountable many complex energy-momentum tensors, $T^{\mu\nu}$ becomes completely real and equal to $T_{\mu\nu}$ every one-fourth the period of the fundamental mode of oscillation; that is whenever $t = \dfrac{nT}{4}$.

Countable symmetry together with Noether's Theorem suggests the energy momentum tensor is conserved under time rotation.

## IV  Redshifted Electromagnetic Radiation.

In this section the metric (2.5) is modified with constant radiant energy ρ (described in terms of the electric and magnetic field strengths). The metric is adjusted slightly to include the radiant energy:

$$\omega \rightarrow \omega + \rho \qquad (4.1)$$

Here the constant value ρ is the radiant energy number density per time; the remainder of the energy units come from the two coefficients of the energy momentum tensor: $(c^2/8\pi G)$. Note that both ω and ρ, when multiplied by time, form the unitless metric:

$$g_{\mu\nu} = e^{i(\omega+\rho)t} \eta_{\mu\nu} \qquad (4.2)$$

On two accounts this amended conformal metric (representing gravitational rotation and electromagnetic radiant energy in the natural exponent) is physically congruous with the perfect fluid radiation dominated universe described by the energy momentum tensor, equation (3.2). Secondly, as is well accepted, the radiant energy density should be three times the radiation pressure[13]. This turns out to be the case but there is also a coupled-gravity-radiant energy term along the diagonal of the tensor.

It follows immediately by the replacement of ω → (ω + ρ), the energy momentum tensor will have the form of:

$$T_{\mu\nu} = \left[\frac{1}{2}I\omega^2 + 2\omega\rho I + I\rho^2\right] \tag{4.3}$$

where I is the same matrix defined by (3.3) [Note that $\omega^2$ has units of $s^{-2}$ as does $\omega\rho$ and $\rho^2$ (radiant energy number squared). When these units are combined with the other units found in I → $c^2/8\pi G$, the resulting units, for all terms in the energy momentum tensor, become Joules/m$^3$].

But how do we interpret equation (4.3)? First, because spacetime is undergoing rotation it is expected that frame dragging occurs (known as the Lense-Thirring effect[14]). It is further assumed and shown mathematically to be the case, that radiant does indeed experience frame dragging.

With this in mind, the first term of the energy momentum tensor (4.3) is interpreted. Its form immediately reveals it to be the rotational kinetic energy of spacetime, itself. The second term, by examination, represents the coupling between the radiant energy $\rho$ and the gravitational field rotating with frequency $\omega$. It is this coupling term that causes the photon to experience frame dragging. The last term represents uncoupled radiant energy propagating through the gravitational field. By conservation of energy, the sum of the second and third term must equal the energy of the radiant prior to entering spacetime rotation. In the next section, this information will be used to calculate the redshift of electromagnetic radiation. And then by reducing the volume of spacetime, to determine the redshift frequency of in an incoming photon.

## V. Measuring the Energy in the Energy Momentum Tensor

What the energy momentum tensor (4.3) provides, is a relationship between radiant energy density and the energy of rotating gravitational system. To be able to measure changes in radiant energy precisely, requires that the energy per volume be made discrete. This can be achieved by shrinking the volume of spacetime down small enough, so that the energy described becomes a Planck packet of discrete energy. It is then necessary to convert the classical energies of rotation and radiation in terms of natural numbers and Planck's constant.

The first step in this conversion is to apply a natural number coefficient [j] to rotational energy term:

$$T_{\mu\nu} = \left[ j\left(\frac{1}{2}I\omega^2\right) + 2r\omega\rho I + I\rho^2 \right] \tag{5.1}$$

and also the real number "r" to calibrates the energy system.

Next the rotational kinetic and radiant energy densities are made discrete by shrinking the volume of spacetime down to the microscopic level. At this level the energies can be though of as packets of discrete energy. This allows for the conversion $I\omega\rho \to h(\omega\omega')^{1/2}/2\pi$, and $I\rho^2 \to h\omega'/2\pi$. Finally, the angular kinetic energy is made discrete by letting: $I\omega^2 \to j$; where j takes on units of angular kinetic energy per volume. With these adjustments, the time component for the energy momentum tensor becomes:

$$\frac{E_{00}}{volume} = \left[ -\frac{1}{2}\frac{j}{volume} - \frac{\hbar}{volume}\left(r\sqrt{\omega\omega'} + \omega'\right) \right] \tag{5.2}$$

The 3/2 of the time component of the I matrix has been absorbed into the volume for reasons of compactness of notation. [What is interesting about this energy relation, is that the numbers j and r, when made very small, describe microscopic energy; but when made large describe macroscopic energy.] Since the volumes are equal, equation (5.2) reduces to:

$$E_{00} = \left[-\left(\frac{1}{2}\right)j - \hbar\left(r\sqrt{\omega\omega'} - \omega'\right)\right] \tag{5.3}$$

The later two terms are of particular interest because by conservation of energy, the sum of these two terms must equal the total energy of the incoming photon before entering into the rotating gravitational field. Therefore:

$$h\nu_{photon} = h\left(\frac{r}{2\pi}\sqrt{\omega\omega'} + \frac{\omega'}{2\pi}\right) \tag{5.4}$$

Where $\omega'/2\pi$ is the uncoupled-redshifted photon frequency. Dividing out Planck's constant and rewriting (5.4), with $\nu_{photon} \rightarrow \nu_0$, the changing redshifted frequency can be determined by the following relation:

$$\nu' = \nu_0 - r\sqrt{\nu_0 \nu_g} \tag{5.5}$$

where $\nu_g$ is the frequency of the gravitational field: $\nu_g = 2.7 \times 10^{-12}$ Hz and $\nu_0$ the frequency of the photon of interest before it interacts with the gravitational field. The reason for the real number r is two fold: First as the photon traverses further into the rotating gravitational field it, an increase in redshift will occur. This can be accounted for by the real number r. Secondly, r can be used to calibrate the redshift equation with known distances and redshifts. To analyze equation (5.5), we must allow r to include the up down spin associated with spacetime rotation. Also, since the further the photon traverses through the gravitational field, the greater the redshift occurs as indicated by (5.5), it is assumed ± r is related directly to radial distance traversed by the photon. The plus and minus is to account for the direction of rotation.

$$\nu' = \nu_0 \pm r\sqrt{\nu_0 \nu_g} \tag{5.6}$$

Depending if the photon is traversing + or – rotational spacetime, equation (5.6) is the equation of a straight line showing a decrease in frequency. The y-intercept is the photon's initial frequency $v_0$ (before entering the rotating gravitational field). The slope is and equal to $[\pm (v_0 v_g)^{1/2}]$ and depends of each initial frequency of the photon.

**VI. Particular Initial Frequencies.**

The first frequency to explore is the most dominant frequency in the universe: that of the back ground radiation frequency. As it turns out, the graviton oscillation frequency ($2.7 \times 10^{-12 \pm 3}$ Hz), times the **2.7 K** cosmic background radiation frequency ($2.79 \times 10^{11}$ Hz) is basically unity. Since the gravitational frequency had a range of a factor of plus or minus three[15], and in view the values of the gravitational frequency and the cosmic background radiation (both cosmically connected), it is assumed the multiplication of the two frequencies is unity:

$$\sqrt{v_g v_{CMBR}} \equiv 1 Hz \qquad (6.1)$$

Where $v_{CMBR}$ is the cosmic background radiation. Therefore equation (5.6), for background radiation becomes:

$$v_{CMBR}' = (2.79 \times 10^{12} \pm r) Hz \qquad (6.2)$$

The greater the real number (corresponding photon path-length), the greater the background radiation redshifts. However, it is conjectured the photon enters the + rotation, the two effects will cancel at the microscopic level of photons interacting with gravitons. The random orientation of graviton rotation means a photon from the cosmic background radiation must undergo random walk of ± energy exchanges with the local rotating gravitons, and thus causing widespread fluctuations in the background radiation[16] that are polarized into E and B-mode patterns[17].

The next interesting photon frequency to examine is the 21-centimeter spectral line created by changes in the energy state of neutral hydrogen. This frequency is 1.420 $x10^9$ Hz. The redshifted frequency is given by:

$$v' = v_0 - n\sqrt{v_0 v_g} = 1.420 x 10^9 \, Hz - r\sqrt{(2.7 x 10^{-12})(1.42 x 10^9) s^{-2}} = \quad (6.3)$$

$$1.420 x 10^9 \, Hz - r 0.062$$

The real number r (corresponding to the distance traversed by the photon through rotational spacetime) must be on the order of $10^9$ to cause any noticeable change in the redshift frequency. Equation (4.15) then becomes:

$$v' = (1.420 - 0.062) x 10^9 \, Hz = 1.358 x 10^9 \, Hz \quad (6.4)$$

Next, it is of interest how far the 21 cm photon must traverse through a rotational gravitational field traversed in field, to obtain the particular redshifted frequency calculated about.

$$d = \frac{c}{r v_0} = \frac{2.99 x 10^8}{(1.42 x 10^9) x 10^9} = 2.1 x 10^{-10} \, meters \quad (6.5)$$

Which surprising is the small size of an atom. However this short distance will tend to be cancelled out by the statistical up-down energy-state of the rotating gravitational field. This means, it might actually take enormous distances for the photon to be statistically redshifted[18,19] distribution function corresponding to redshift and random walk distance.

**VII. Conclusion**

It was found if the spacetime metric is extended to include constant radiant energy: $g_{\mu\nu} = e^{i(\omega+\rho)t} \eta_{\mu\nu}$, then the energy momentum tensor is given by: $T_{\mu\nu} = [½ I\omega^2 + (3c^2/16G)(\rho^2 + 2n\omega\rho)]$. In a local rotational gravitational field, this radiant energy was reduced to a single photon with initial frequency $\nu_0$. This General Relativistic result shows as a photon traverses through the rotating gravitational field, it becomes coupled to the field and so redshifts by the amount $\nu' = \nu_0 - r(\nu_0 \nu_g)^{1/2}$.

---

[1] W. J. Christensen Jr. "Normal Coordinates describing coupled oscillations in the gravitational field." Gen. Rel. Grav. 39:105-110 (2007) DOI: 10.1007/s10714-006-0360-8

[2] The German mathematician Karl Weierstrass (1815-1897) showed in 1858 that the motion of a dynamical system could always be expressed in terms of normal coordinates.

[3] W. Misner, K. S. Thorne and J. Wheeler. Gravitation. W. H. Freeman and Company. 1973. Pages 1022-1034

[4] N. Byers. Israel Mathematical Conference Proceedings Vol. 12, (1999)

[5] H. Goldstein, C. Poole, J. Safko. Classical Mechanics, Addison Wesley. 250 (2002)

[6] J. B. Marion and S. T. Thornton. Classical Dynamics of Particles and Systems, HBJ. 458 (1988, 3rd edition)

[7] R. M. Quick and H. G. Miller. Phys. Rev. D **31,** 2682 (1985)

[8] A. Majid, A. Allezy, R. Dufour. Journal of Vibrations and Acoustics. **128,** 50 (2006)

[9] H. Collins, B. Holdom. Phys. Rev. D **65,** 124014-1 (2002)

[10] Private Communication with Prof. Alfonso Agnew, of California State University Fullerton, Mathematics Department independently confirmed the energy-momentum tensor result of equation (3.1)